\documentclass[twocolumn,showpacs,preprintnumbers,amsmath,amssymb]{revtex4}
\usepackage{graphicx}% Include figure files
\usepackage{bm}% bold math
\begin{document}
\preprint{...}
\title{Magnetomotive drive and detection of clamped-clamped mechanical resonators in water}
\author{W.J. Venstra}
\author{H.J.R. Westra}
\author{K. Babaei Gavan}
\author{H.S.J. van der Zant}
\affiliation{Kavli Institute of Nanoscience, Delft University of Technology, Lorentzweg 1, 2628CJ Delft, The Netherlands}%
\date{December 30, 2009}
\begin{abstract}
We demonstrate magnetomotive drive and detection of doubly clamped
string resonators in water. A compact 1.9 T permanent magnet is
used to detect the fundamental and higher flexural modes of
$\mathrm{200\,\mu m}$ long resonators. Good agreement is found
between the magnetomotive measurements and optical measurements
performed on the same resonator. The magnetomotive detection
scheme can be used to simultaneously drive and detect multiple
sensors or scanning probes in viscous fluids without alignment of
detector beams.
\end{abstract}
 \maketitle
\indent\indent Micro- and nanomechanical resonators have many
applications as scanning probes and mass or stress sensors.
Several techniques have been developed to detect resonator
vibrations in vacuum and at atmospheric pressure. The natural
environment for biological experiments however is an aqueous
solution, and detection methods in liquid environments are less
numerous. Optical or piezoresistive schemes are commonly used,
combined with a separate excitation source, e.g.
magnetic~\cite{Vancura05} or piezoelectric, to drive the strongly
damped resonator.\\
\indent\indent In this work we demonstrate that a magnetomotive
technique can be used to drive and detect micromechanical
resonator vibrations in water. The magnetomotive technique allows
straightforward resonator geometry, a strong driving force acting
directly on the resonator, and it can be scaled towards nanometer
dimensions. The technique has been applied in vacuum to
characterize resonators up to the $\mathrm{GHz}$-range
~\cite{Huang03}, in mass sensing~\cite{Huang05}, and to readout
resonator arrays driven in the nonlinear regime at atmospheric
pressure~\cite{Venstra08}.\\

\begin{figure}[b]
\includegraphics[width=85mm]{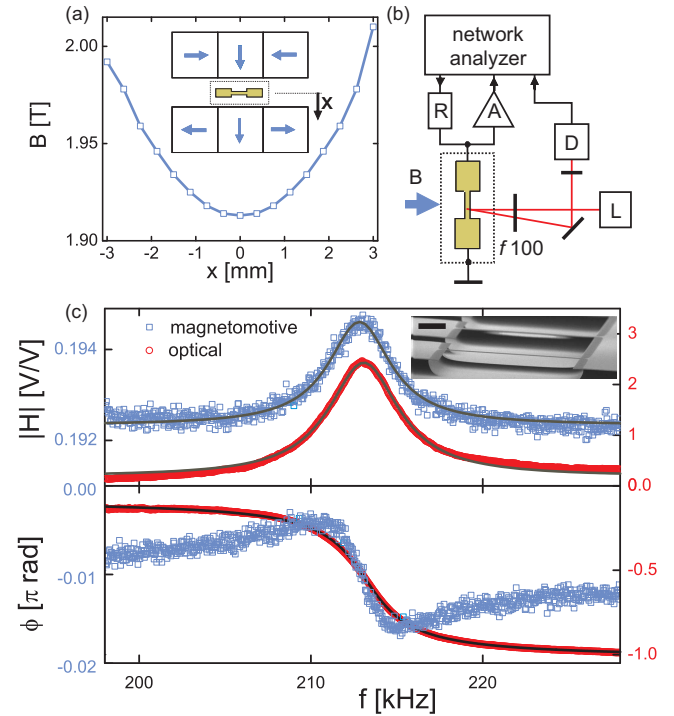}
\caption{(a) Magnetic field strength as a function of the position
in the gap between the magnet poles, $x$. Inset shows a
cross-section of the magnet construction, squares represent the
individual magnets, polarized according to the arrows. (b)
Schematic of the combined magnetomotive and optical setup. (c)
Frequency response functions of the magnetomotive driven resonator
in air, as measured by magnetomotive and optical techniques,
magnitude $\mathrm{|H|}$ and phase $\mathrm{\phi}$, the black
solid lines represent driven harmonic oscillator functions fits.
Inset: scanning electron micrograph of two clamped-clamped string
resonators, scale bar is $\mathrm{20\,\mu m}$.\label{setup}}
\end{figure}
\indent\indent A compact and powerful permanent magnet is
constructed to drive the strongly damped resonator in the fluid.
The magnet is composed of commercially available NdFeB magnets
with a remanent induction of $\mathrm{B_r=1.5\,T}$ and a total
volume of $\mathrm{213\,cm^3}$. By configuring the magnets in a
Halbach array~\cite{Halbach83}, a field strength up to
$\mathrm{2\,T}$ is generated in a $\mathrm{6\,mm}$ gap between the
magnet poles. Figure 1(a) shows the measured field as a function
of the position in the gap, $x$. The inset shows a schematized
cross section of the construction, the arrows indicate the
polarization of the magnets. The field strength varies less than
$5\%$ within a volume of $\mathrm{6\times 6\times 6\,mm^3}$ inside
the gap, and is minimum in the center at $x\,\mathrm{=0}$,
where the resonator is located. \\
\indent\indent Figure 1(b) shows the measurement setup. The
alternating voltage from a network analyzer is applied to the
resonator via a series resistor, ${\mathrm{R=25 \, \Omega}}$. The
generated electromotive force is measured via a high impedance
buffer, marked A in the figure, on one input channel of the
analyzer. As a reference, the deflection of the resonator is
probed at the same time by a Helium Neon laser, L. The beam
reflection is captured on a linear position sensitive detector, D,
and measured on the second input channel of the network analyzer.\\
\indent\indent The resonator is placed in a custom-built
temperature controlled flow cell with a volume of $\mathrm{3\,\mu
l}$, such that the Lorentz force is directed out-of-plane,
corresponding to the fundamental resonance mode.\\
\indent\indent The resonators are fabricated from
$\mathrm{100\,nm}$ thick low-pressure chemical vapor deposited
silicon nitride using electron beam lithography and reactive ion
etching, and suspended using a dry isotropic release process. The
inset in Fig. 1(c) shows two resonators before metallization. The
dimensions are $L \times w = \mathrm{200\times 15\,\mu m^2}$. A
$\mathrm{5\,nm}$ thick chromium adhesion layer is deposited on
top, followed by $\mathrm{30\,nm}$ of gold. The gold layer is not
passivated and is in contact with the water
during the experiments.\\
\indent\indent In air, the fundamental resonance mode, measured by
the magnetomotive and optical techniques is shown in Fig.1(c). The
signal to noise ratio for the optical measurement is $\mathrm{50}$
times larger, which is due to the approximately $200$ times
amplification of the resonator displacement by the optical lever.
The resonator resistance, $\mathrm{R \approx7.5\, \Omega}$, is
large compared to its motional impedance,
$\mathrm{Z_0\approx0.8\,\Omega}$ at resonance, and this results in
a phase response for the magnetomotive measurement different from
the driven harmonic oscillator response obtained by the optical
detector. By fitting damped driven harmonic oscillator functions
represented by the solid black lines, the natural (undamped)
resonance frequency for the fundamental mode is
$f_\mathrm{1,air}\mathrm{=213\,kHz}$, and the quality factor
$Q_\mathrm{1,air}=50$. The magnetomotive technique can be applied
to higher odd modes, though the sensitivity reduces, as will be
discussed later. The third resonance mode is also measured with
$f_\mathrm{3,air}\mathrm{=725\,kHz}$. For the n-th mode of
vibration, the frequency ratio $f_\mathrm{n}/f_\mathrm{1}=n$ for a
string in tension, and roughly $(\frac{2n+1}{3})^2$ for a doubly
clamped flexural beam \cite{Weaver}. The measured ratio
$f_\mathrm{3,air} / f_\mathrm{3,air} = 3.4$, indicates string-like
behavior dominates. This occurs when the restoring force from
flexure is small compared to that from residual tension. We note
that the ratio is $\mathrm{13\%}$ larger than for a string in
tension, and this is explained by the flexural rigidity which
contributes to some increase in effective spring stiffness,
whereas the compliance from the large undercuts
decreases it slightly.\\

\begin{figure}[t]
\includegraphics[width=85mm]{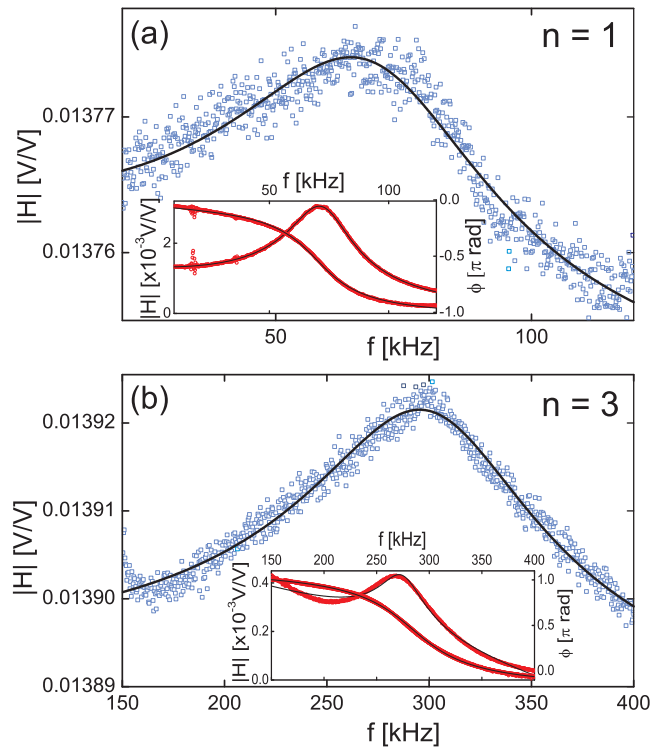}
\caption{(a) Magnetomotive measurement of the fundamental
resonance mode of the string resonator in water. The inset shows
the optical measurement. (b) Magnetomotive and optical measurement
of the third flexural mode.}
\end{figure}
\indent\indent Prior to experiments in water, the flow cell is
flushed with ethanol to remove air bubbles. De-ionized water is
injected and the optical and magnetomotive measurements are
repeated on an immersed resonator. Figure 2(a) shows the measured
amplitude response using the magnetomotive scheme for the first
resonance mode. A harmonic oscillator function is fit through the
data with $f_\mathrm{1,water}\mathrm{=74\, kHz}$, which is
identified as a mechanical resonance by the magnitude and phase
response from the optical measurement, shown in the inset. A
different Q-factor is found for the magnetomotive and optical
detector. The magnetomotive quality factor,
$Q_\mathrm{1,water}\mathrm{=1.4}$ is lower than for the optical
measurement $Q_\mathrm{1,water}\mathrm{=2.8}$. This difference is
attributed to the complex dielectric constant of water, which
presents a frequency-dependent load to the resonator and lowers
the Q-factor for the magnetomotive measurement \cite{Cleland99}.
The value of the load depends on the design of sample and flow
cell and on the frequency-dependent dielectric properties of
water, and is not further discussed here.\\
\indent\indent The third flexural mode of the same resonator is
plotted in Fig. 2(b). The magnetomotive detected resonance
frequency equals $f_\mathrm{3,water}\mathrm{=306 \,kHz}$ and
$Q_\mathrm{3,water}=2.7$. The optical measurement reveals a
slightly lower resonance frequency,
$f_\mathrm{3,water}\mathrm{=275 \,kHz}$. This difference was found
independent on sweep time and driving strength, and cannot be
explained by electrical loading of the resonator. The optically
measured Q-factor equals $Q_\mathrm{3,water}= 6.4$, again higher
than the magnetomotive one. In water the ratio between the
resonance frequencies is similar to the ratio in air:
$f_\mathrm{3,water}/f_\mathrm{1,water} = 3.8$.\\
\indent\indent To compare the shift in resonance frequency and
Q-factor upon immersion with theory, we take the viscous and
inertial forces into account through a hydrodynamic function,
$\Gamma(\mathrm{Re,\kappa})$, which relates the cross section
shape of the resonator to a force per unit length acting on the
resonator, where its real and imaginary components correspond to
the inertial and dissipative components of the force
\cite{Sader98}. $\Gamma(\mathrm{Re,\kappa)}$ is a function of the
normalized mode-dependent Reynolds number, $\mathrm{Re_n}=2\pi
f_\mathrm{n,vac}\rho /\eta$, and the normalized mode number,
$\mathrm{\kappa_n=\pi n \frac{w}{L}}$ for our string resonators.
Here $\mathrm{\rho}$ is the density and $\mathrm{\eta}$ the
dynamic viscosity of water. The hydrodynamic function for a
rectangular cross section is given in Ref.~\cite{Sader07}, and we
can calculate the ratio between the frequencies:
\begin{equation}
\frac{f_\mathrm{n,water}}{f_\mathrm{n,vac}}=[1+\frac{\pi w \rho}{4
t \rho_\mathrm{c}}\Gamma_\mathrm{R}(\mathrm{Re,\kappa)}]^{-0.5},
\end{equation}
where $\rho_\mathrm{c}$ denotes the average density of the gold
coated resonator,  $\mathrm{\Gamma_R}$ denotes the real part of
$\Gamma(\mathrm{Re,\kappa})$, and $\mathrm{\pi w \rho /4t\rho_c}$
represents the mass loading parameter \cite{Villa09}. Although
this model assumes a high Q--factor, experiments have shown it
accurately describes the response of cantilever beams with
Q--factors close to ours \cite{Sader00}. The results for the
resonance frequencies and Q-factors are summarized in Table 1. We
measured the resonance frequency from thermal noise spectra in
atmospheric pressure and in the intrinsic damping regime in
vacuum, and found $f_\mathrm{1,air}/f_\mathrm{1,vac}=1.013$ and
$f_\mathrm{3,air}/f_\mathrm{3,vac}=1.007$, which is negligible
compared to the shifts upon immersion in water \cite{Qairvac}. We
can thus assume $f_\mathrm{n,air} = f_\mathrm{n,vac}$, and the
observed ratio's $f_\mathrm{n,water}/f_\mathrm{n,vac}$ are close
to the theoretical prediction. The slightly higher prediction of
the resonance frequency and Q-factors may be explained by the
limited distance between the resonator and the substrate. In our
experiment the distance is comparable to the resonator width, and
in this case additional friction reduces the Q-factor by a
factor of 2 when compared to a freely moving resonator \cite{Gittes96}.\\
\begin{table}
\caption{Resonance frequency and Q-factor of mode $n$ in air and
water, measured on a second device. The theoretic and experimental
ratio's between the resonance frequencies upon immersion are
$r_\mathrm{theory}=f_\mathrm{water}/f_\mathrm{vac}$ and
$r_\mathrm{exp}=f_\mathrm{water}/f_\mathrm{air}$.}
\begin{ruledtabular}
\begin{tabular}{cccccccccc}
&\multicolumn{5}{c}{frequency (kHz)}&\multicolumn{4}{c}{$\mathrm{Q-factor}$}\\
n& \vline %%
&air&water& $r_\mathrm{theory}$& $r_\mathrm{exp}$&\vline &air&water&water,theory\\
\hline 1&\vline& 194.9  & 52.6  &  0.23 & 0.27&\vline&50 &2.0&4.1\\
       3&\vline & 593.3  & 184   & 0.25  &  0.31&\vline&67&5.4&6.7\\
\end{tabular}
\end{ruledtabular}
\end{table}
\indent\indent To investigate the possibility to scale-down the
dimensions of the string resonators, we have calculated the
displacements and electromotive voltages at
resonance~\cite{Yurke95}. In the harmonic regime, the tension
force $T$ is constant and sufficient to ensure string-like
behavior with mode shapes and amplitudes $\mathrm{N_n}$ given by
$\mathrm{u_n(x)=N_n\sin(\pi nx/L)}$. The maximum displacement of
the string for $\mathrm{n}=1,3,..$, can then be calculated as
\cite{derivationEq2}:
\begin{equation}
u_\mathrm{max,n}=\frac{4BIQ_nL^2}{n^3\pi^3T}
\end{equation}
with $\mathrm{B}$ the magnetic field and $I$ the current through
the resonator. For currents in the mA-range and an estimated
residual stress of $\mathrm{60\,MPa}$, the amplitude at resonance
is about $\mathrm{u_{max,1}=10\,nm}$ for the first mode, and
decreases with the mode number cubed. We note that the excellent
thermalization in water allows large drive currents without
damaging the resonator, and large resonator amplitudes can be
obtained. The generated electromotive voltage is given by
\begin{equation}
V_{\mathrm{EMF,n}}=\frac{8}{\pi^3}\frac{f_nI B^2Q_nL^3}{n^3T}
\label{eq3}
\end{equation}
for odd mode numbers. For even $n$, the net flux rate is zero and
no voltage is generated. The resonance frequency $f_\mathrm{n}$
scales with the inverse dimension ($d^{-1}$). Furthermore, in
these experiments the tension force, $\mathrm{T}$, results from
residual stress and therefore scales as ($wt=d^{2}$). In a
harmonic spectrum, $f_\mathrm{n}=\mathrm{n}f_\mathrm{1}$, the
generated voltage is thus independent on the mode number and
proportional to the displacement $u_\mathrm{n,max}$, which is
adjustable by the driving force. Therefore {$V_{\mathrm{EMF}} \sim
fL^3/T  \sim d^0$}, independent on dimension.\\

\begin{figure}[t]
\includegraphics[width=85mm]{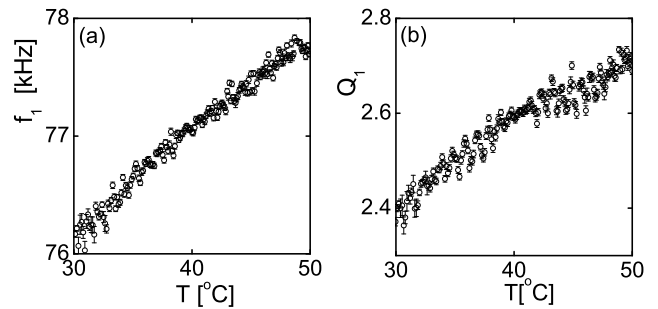}
\caption{Temperature dependence of the resonance frequency (a) and
Q-factor (b) of the fundamental resonance mode of an immersed
string resonator.}
\end{figure}

\indent\indent Changes in liquid density and viscosity associated
with temperature fluctuations pose a limit on the stability and
accuracy of frequency-based measurements. Electrical and
mechanical energy is dissipated by the strongly driven resonators
(i.e. by ohmic resistance and by the viscous force), and an
increase of water temperature is the result. The electrical and
mechanical dissipation can be expressed as $
U_\mathrm{in}=I^2R+2\pi^2 m f^3_nu_{max}^2/Q$. In our experiments
the contributions are on the order of
$\mathrm{U_{in,e}=10^{-5}\,J/s}$ for the electrical and
$\mathrm{U_{in,m}=10^{-12}\,J/s}$ for the mechanical dissipation,
and electrical dissipation is dominant. If we apply the input
energy to heat the $\mathrm{3\,\mu l}$ volume of water inside the
flow cell, a temperature increase on the order of
$\mathrm{60\,mK/s}$ is the result. We investigated the effect of
such temperature fluctuations by measuring the resonance frequency
and Q-factor during a controlled temperature sweep over a range of
$\mathrm{20\,^oC}$. As shown in Fig. 3, both the resonance
frequency and the Q-factor increase while increasing the
temperature, as expected due to the reduction of the fluid density
and viscosity. The results presented here for doubly clamped
string resonators are comparable to results obtained earlier for
cantilever beams~\cite{Kim06}. The temperature dependence of the
resonance frequency, approximately $\mathrm{100\,Hz/^oC}$ in this
experiment, underlines the need for accurate temperature control
during mass sensing experiments in liquids.\\
\indent\indent In summary, we demonstrated magnetomotive drive and
detection of micromechanical string resonators in water. The
fundamental and third flexural resonance modes were detected, and
the observed changes in resonance frequency and Q-factor are in
agreement with theory. The magnetomotive technique can be applied
to obtain large vibration amplitudes in water, and to detect the
resonator motion without alignment of probe beams, as with optical
detection schemes. Higher modes can be used to enhance
sensitivity, and to measure the mass of single particles
regardless of the location of the accretion on the resonator
surface~\cite{Dohn06}.\\
\indent\indent The authors acknowledge financial support from the
Dutch organizations FOM and NWO (VICI), and Koninklijke Philips NV
(RWC-061-JR-05028).
\newpage


\begin{thebibliography}{18}
\bibitem[{Qai()}]{Qairvac}
\bibinfo{note}{The ratio's of the Q-factors are $Q_\mathrm{1,air}/Q_\mathrm{1,vac}\approx0.12$ and
  $Q_\mathrm{3,air}/Q_\mathrm{3,vac}\approx0.31$.}
\bibitem[{der()}]{derivationEq2}
\bibinfo{note}{For the tension dominated regime
  $u_\mathrm{max,n}=\frac{BIQ_\mathrm{n}N_\mathrm{n}\int^{L}_{0}
  u_\mathrm{n}(x)dx}{T\int^{L}_{0}(\frac{du_\mathrm{n}(x)}{\partial x})^2\partial x}$,
  where the modeshapes are given by $\mathrm{u_n(x)=N_n\sin(\pi nx/L)}$.}
\end{thebibliography}
\end{document}